\documentclass[a4paper,11pt]{article}
\usepackage{graphicx}

\usepackage{multicol}
\usepackage{color}
\definecolor{rosso}{cmyk}{0,1,1,0.4}
\definecolor{rossos}{cmyk}{0,1,1,0.55}
\definecolor{rossoc}{cmyk}{0,1,1,0.2}
\definecolor{blu}{cmyk}{1,1,0,0.3}
\definecolor{blus}{cmyk}{1,1,0,0.6}
\definecolor{bluc}{cmyk}{1,1,0,0.1}
\definecolor{verde}{cmyk}{0.92,0,0.59,0.25}
\definecolor{verdec}{cmyk}{0.92,0,0.59,0.15}
\definecolor{verdes}{cmyk}{0.92,0,0.59,0.4}
\newcommand{\beq}{\begin{equation}}
\newcommand{\eeq}{\end{equation}}

\newcommand{\GeV}{\,{\rm GeV}}
\newcommand{\TeV}{\,{\rm TeV}}

\newcommand{\SU}{\,{\rm SU}}

\def\Lag{\mathscr{L}}
\def\Lag{{\cal L}}

\newcommand{\mT}{\mb{m}_T}

\advance\textheight by \topskip
\large\normalsize
\newcommand{\be}{\begin{equation}}
\newcommand{\ee}{\end{equation}}
\newcommand{\bea}{\begin{eqnarray}}
\newcommand{\eea}{\end{eqnarray}}
\newcommand{\mrm}[1]{\mbox{\rm #1}}
\newcommand{\Eq}[1]{eq.~(\ref{#1})}
\newcommand{\rfn}[1]{(\ref{#1})}
\newcommand{\ba}{\begin{array}}
\newcommand{\ea}{\end{array}}

\newcommand{\eV}{\,{\rm eV}}

\newcommand{\eq}[1]{~(\ref{eq:#1})}

\newcommand{\PL}{Phys. Lett.}
\newcommand{\PR}{Phys. Rev.}
\newcommand{\mb}[1]{\mbox{\normalsize\boldmath $#1$}}
\newcommand{\fig}[1]{~\ref{fig:#1}}

\def\circa#1{\,\raise.3ex\hbox{$#1$\kern-.75em\lower1ex\hbox{$\sim$}}\,}
\makeatletter
\def\art{\@ifnextchar[{\eart}{\oart}}
\def\eart[#1]#2#3#4#5#6{{\rm #2}, {#3 #4} {\rm (#6) #5} ({#1})}
\def\hepart[#1]#2{{\rm #2, #1}}
\newcommand{\oart}[5]{{\rm #1}, {#2 #3} {\rm (#5) #4}}

\newcounter{alphaequation}[equation]
\def\thealphaequation{\theequation\hbox to
0.6em{\hfil\alph{alphaequation}\hfil}}
\def\eqnsystem#1{
\def\@eqnnum{{\rm (\thealphaequation)}}
\def\@@eqncr{\let\@tempa\relax \ifcase\@eqcnt \def\@tempa{& & &} \or
   \def\@tempa{& &}\or \def\@tempa{&}\fi\@tempa
   \if@eqnsw\@eqnnum\refstepcounter{alphaequation}\fi
\global\@eqnswtrue\global\@eqcnt=0\cr}
\refstepcounter{equation} \let\@currentlabel\theequation \def\@tempb{#1}
\ifx\@tempb\empty\else\label{#1}\fi
\refstepcounter{alphaequation}
\let\@currentlabel\thealphaequation
\global\@eqnswtrue\global\@eqcnt=0 \tabskip\@centering\let\\=\@eqncr
$$\halign to \displaywidth\bgroup \@eqnsel\hskip\@centering
$\displaystyle\tabskip\z@{##}$&\global\@eqcnt\@ne
\hskip2\arraycolsep\hfil${##}$\hfil&  
\global\@eqcnt\tw@\hskip2\arraycolsep
$\displaystyle\tabskip\z@{##}$\hfil
\tabskip\@centering&\llap{##}\tabskip\z@\cr}
\def\endeqnsystem{\@@eqncr\egroup$$\global\@ignoretrue} \makeatother

\begin{document}
\centerline{\hfill IFUP--TH/2005-17}
\color{black}
\vspace{1.0cm}
\centerline{\LARGE\bf\color{rossos}Efficiency and maximal CP-asymmetry}
\centerline{\LARGE\bf\color{rossos}of scalar triplet leptogenesis}
\medskip\bigskip\color{black}\vspace{0.6cm}
\centerline{\large\bf Thomas Hambye$^a$,  Martti Raidal$^b$,
Alessandro  Strumia$^c$}\vspace{0.5cm}
\centerline{\em $^a$ Department of Physics, Theoretical Physics,
University of Oxford, Oxford OX1\hspace{0.2em}3NP, UK }\vspace{0.13cm}
\centerline{\em $^b$ National Institute of Chemical Physics and Biophysics,
Ravala 10,Tallinn 10143, Estonia}\vspace{0.13cm}
 \centerline{\em $^c$ Dipartimento di Fisica dell'Universit\`a di Pisa
and INFN, Italia}\vspace{1.cm}

\centerline{\large\bf\color{blus} Abstract}
\begin{quote}
\vspace{-0.25cm}
\indent\color{blus}\large
We study thermal leptogenesis induced by 
decays of a scalar $\SU(2)_L$ triplet.
Despite the presence
of gauge interactions,
unexpected features of the Boltzmann equations
make the efficiency close to maximal
in most of the parameter space. 
We derive the maximal CP asymmetry
in triplet decays, assuming that it is generated by
heavier sources of neutrino masses: in this case
successful leptogenesis needs a triplet heavier than 
$2.8 \cdot 10^{10}$~GeV 
and does not further restrict its couplings, 
allowing detectable $\mu\to e\gamma,\tau\to\mu\gamma$ rates 
in the context of supersymmetric models.
Triplet masses down to the TeV scale are viable
in presence of extra sources of CP-violation.

\color{black}
\end{quote}

\section{Introduction}
Majorana neutrino masses can be mediated by tree-level exchange of
three different kinds of new particles:
I) fermion singlets~\cite{type-I};
II) scalar  SU(2)$_L$ triplets~\cite{type-II};
III) fermion SU(2)$_L$ triplets~\cite{tripletferm}.
Indeed these particles can have renormalizable
couplings to lepton doublets $L$ and Higgs $H$,
generating the unique Majorana neutrino mass operator $(LH)^2$.
Low energy experiments can see neutrino masses and
reconstruct the coefficients of the $(LH)^2$ operator,
but cannot tell their origin.
More information about the unknown high-energy theory that generates neutrino masses
can be obtained assuming that the observed baryon asymmetry of the Universe
is produced via thermal leptogenesis~\cite{FY} in decays of the lightest
of these three kinds of particles.

Leptogenesis has been extensively studied only in case I)~\cite{FY,GNRRS}.
The main concern with the other possibilities is that 
gauge scatterings can keep
$\SU(2)_L $ triplets close to thermal equilibrium,
conflicting with the third Sakharov condition~\cite{Sakharov}.
Since gauge scatterings are slower than the expansion rate of the Universe
only at temperatures $T\circa{>} 10^{15}$~GeV~\cite{Olive,sasa},
it was generally expected that 
successful triplet leptogenesis is possible only
around that energy scale. 
However, first estimates of the leptogenesis efficiency in scalar 
triplet decays~\cite{scaltriplepto,hs,seesaw25TH}
as well as the full calculation for fermion triplet~\cite{HLNPS} have shown 
that thermal leptogenesis is efficient enough even at lower temperatures.


In this paper we present the first full calculation of thermal 
leptogenesis in 
the decays of a $\SU(2)_L $ triplet scalar $T$. 
We start deriving the maximal  CP asymmetry $\varepsilon_L$ 
in triplet  decays generated by any other source of neutrino masses
much heavier than $T$  (e.g.\ additional triplets $T'$, right-handed 
neutrinos, etc).
We next derive and solve the full set of Boltzmann equations describing the
thermal evolution of the relevant abundances.
We find that thermal leptogenesis from scalar triplets proceeds in a
qualitatively different way from the alternative scenarios studied so far.
In particular, a quasi-maximal efficiency can be obtained even if
gauge scatterings 
keep the triplet abundancy very close to thermal equilibrium.


The CP asymmetry induced by the heavier sources of 
neutrino masses decreases with the triplet mass $M_T$,
so that successful  leptogenesis needs 
$
M_T>2.8 \times 10^{10}\;\; \mrm{GeV}$.
More general sources of CP violation allow thermal leptogenesis even down to $M_T\sim\TeV$:
such light triplets can be tested at collider experiments~\cite{HMPR}.

Our paper is organized as follows.
Section~\ref{T} contains the technical details necessary for this analysis.
Results are discussed in section~\ref{res} and summarized in the conclusions.

\section{Scalar triplet}\label{T}

We start by presenting the model under consideration.
The relevant parts of the scalar triplet  Lagrangian are
\beq\label{eq:tripletS}
\Lag = \Lag_{\rm SM} + |D_\mu T|^2- M_T^2 |T^a|^2+\frac{1}{2}
\bigg(\mb{\lambda}^{gg'}_L L^i_g  \tau^a_{ij}  \epsilon L^j_{g'} T^a  + 
M_T\lambda_H\, H^i\tau^a_{ij} 
\epsilon H^j T^{a*}+\hbox{h.c.} \bigg),
\eeq
where $3\times3$ flavour matrices are denoted in bold-face, 
$g,g'=\{1,2,3\}$ are generation indices,
$\epsilon$ is the permutation matrix
and $\tau^a$ are the usual $\SU(2)_L$ Pauli matrices.
The hypercharges are $Y_L=-1/2$, $Y_H=1/2$ and $Y_T=1$.
The triplet Lagrangian can be supersymmetrized introducing two 
chiral superfields $T$ and $\bar{T}$
with superpotential couplings:
\beq\label{eq:tripletSUSY}
W = W_{\rm MSSM} + M_T T\bar{T} + \frac{1}{2}
 \bigg(
\mb{\lambda}^{gg'}_L L_g L_{g'} T  + 
\lambda_{H_{\rm d}}  H_{\rm d} H_{\rm d} T +
\lambda_{H_{\rm u}}  H_{\rm u} H_{\rm u} \bar{T}\bigg).\eeq
Triplet exchange mediates the dimension-5 neutrino mass operator 
$(LH_{\rm u})^2$ such that
the triplet contribution $\mb{m}_T$ to the  neutrino mass matrix $\mb{m}_\nu$ is
\beq 
\mT =  \mb{\lambda}_L \lambda_{H_{\rm u}}\frac{v_{\rm u}^2}{M_T}
\label{mT}
\eeq
where 
$v=174\GeV$ and the subscript u is present only in the SUSY version of the model.

\subsection{Decay rates and CP-asymmetry}
From now on we work explicitly with the non-supersymmetric 
version of the model.  The tree-level triplet decay rates are
\beq\Gamma(T\to LL) =\frac{M_T}{16\pi} 
{\rm Tr}\, \mb{\lambda}_L \mb{\lambda}_L^\dagger=B_L \Gamma_T ,\qquad
\Gamma(T\to \bar H\bar H) =\frac{M_T}{16\pi} 
\lambda_H \lambda_H^\dagger=B_H \Gamma_T ,
\eeq
where $B_L$ and $B_H$ are the tree-level branching ratios to leptons and Higgs
doublets, respectively, and $\Gamma_T$ is the total triplet decay width.
Assuming that these are the only decay modes, i.e. $B_L+B_H=1$,
and taking into account
CPT-invariance,
a single parameter $\varepsilon_L$ determines CP-violation in $T,\bar{T}$ 
decays to be:
\begin{equation}\label{eq:GammaTs}\begin{array}{rclrcl}
\Gamma({\bar{T}}\to LL)  &=& \Gamma_T\,(B_L +{\varepsilon_L}/{2}
), \qquad&
\Gamma({\bar{T}}\to {\bar{H}}{\bar{H}})  &=& \Gamma_T\,(B_H-{\varepsilon_L}/{2}
), \\
\Gamma(T\to \bar{L}\bar{L})  &=& \Gamma_T\,(B_L -{\varepsilon_L}/{2}
), &
\Gamma(T\to HH)  &=&\Gamma_T\,(B_H+{\varepsilon_L}/{2}
),\end{array}
\end{equation}
where $\varepsilon_L$ is the CP-asymmetry (i.e.~the average lepton number 
produced per decay):
\begin{equation}\label{eq:epsLepsT}
\varepsilon_L\equiv 2  \frac{\Gamma({\bar{T}}\rightarrow LL)-
\Gamma(T\rightarrow \bar{L}\bar{L})}{\Gamma_{T}+\Gamma_{\bar T}}.
\end{equation}
The overall factor 2 arises because $\bar T \to LL$ generates 2 leptons.
Defining, as usual, 
the efficiency factor to be unity in the limit where the triplets 
decay strongly out-of-equilibrium, the lepton to photon number density 
ratio produced is
\beq
\frac{n_L}{n_\gamma}= \varepsilon_L \eta  
\left.\frac{n_T+n_{\bar{T}}}{n_\gamma}\right|_{T \gg M_T},
\eeq
where $n_T$ is the total triplet number density 
$n_T=n_{T^{--}}+n_{T^{-}}+n_{T^{0}}$. 
After partial conversion of the lepton asymmetry to a baryon asymmetry 
by sphalerons this leads to
\beq
\label{eq:uusm}
\frac{n_B}{n_\gamma}=-0.029 \varepsilon_L \eta \,.
\eeq 

One triplet $T$ alone can mediate the whole observed light neutrino mass 
matrix via \Eq{mT}
(see e.g.~\cite{Senj} for models of this type). 
However, in this case the 
CP-asymmetry $\varepsilon_L$ is generated only at higher loops and
 is highly suppressed. To get a sizable CP-asymmetry 
 extra couplings are needed.
 The minimal option is that the CP-asymmetry
 is generated by extra contributions to neutrino masses,
 mediated e.g.\ by heavier right-handed neutrinos (fig.~1a \cite{sod,hs}),
 or by additional heavier triplets 
(fig.~1b \cite{masar,scaltriplepto}) or by any other heavier particle
which induces the dimension-5 operator as shown in fig.~1c.
In this case the neutrino mass $\mb{m}_\nu$ is given by the 
sum of the triplet contribution $\mT$,
plus an extra contribution $\mb{m}_H$ mediated by these $H$eavier particles:
$\mb{m}_\nu = \mT + \mb{m}_H$.
\begin{figure}[t]
$$\hspace{-5mm}
\includegraphics[width=0.9\textwidth]{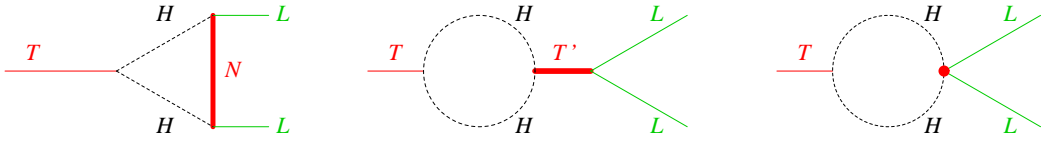}$$
\caption{\label{fig:FeynCP}\em
One-loop diagrams contributing to the asymmetry
in scalar triplet decays.}
\end{figure}
Assuming that these particles 
are substantially heavier than the scalar triplet,
the neutrino mass contribution to the CP-asymmetry is
\beq\label{eq:DIT} 
|\varepsilon_L|  =   \frac{1}{4\pi}\frac{M_T}{v^2}\sqrt{B_L B_H}
 \frac{|{\rm Im}\,{\rm Tr}\, \mT^\dagger \mb{m}_H|}{\tilde{m}_T},\eeq
 where $\tilde{m}_T^2\equiv {\rm Tr}\,\mT^\dagger\mT$. 
We stress that, as for the decay of a right-handed  
neutrino \cite{AK}, this result is independent of the nature of the heavier particle 
contributing to the neutrino masses (fig.~1a or fig.~1b or whatever) 
because  the heavier particle effects on the asymmetries
 can be fully encoded 
in $\mb{m}_H$, i.e.~in term of the unique dimension-5 neutrino mass 
operator they induce 
(fig.~1c).\footnote{In the case of fig.~1a, our result in 
eq.~(\ref{eq:epsmax}) differs by a factor 1/2 from the first 
calculation performed in~\cite{hs}.}

 We remind that when the lightest particle is a right-handed neutrino $N_1$ 
 the CP-asymmetry in its decays is given by an analogous formula~\cite{CP,leptogenesisBounds}:
 \beq\label{eq:DIN} 
 | \varepsilon_1 |  = \frac{3}{16\pi}\frac{M_N}{v^2} 
 \frac{|{\rm Im}\,{\rm Tr} \,{\mb{m}}_1^\dagger \mb{m}_H|}{\tilde{m}_1}
\le
 \frac{3}{16\pi}\frac{M_1}{v^2} (m_{\nu_3} - m_{\nu_1}),
\eeq
where, to make contact with the standard notation,
the contribution to the neutrino mass matrix mediated by $N_1$
has been denoted as $\tilde{\mb{m}}_1$.
The bound of eq.~(\ref{eq:DIN}) holds assuming that the heavier particles 
are two right-handed neutrinos~\cite{leptogenesisBounds}.
 In the case of generic heavy particles $m_{\nu_3}-m_{\nu_1}$ gets replaced by $m_{\nu_3}$.

In a similar way one obtains an upper bound on the 
neutrino mass contribution to the triplet
CP-asymmetry of \Eq{eq:DIT}:
 \beq \label{eq:epsmax}  | \varepsilon_L |  \le 
 \frac{1}{4\pi} \frac{M_T}{v^2}\sqrt{B_L B_H\sum_i  m_{\nu_i}^2} .\eeq
The result is different from the singlet case
because $\mT$ is a generic matrix, while $\tilde{\mb{m}}_1$ has  rank~1.
As a result $\varepsilon_L$ increases 
for larger  $m_{\nu_i}$ \cite{hs}, 
unlike $\varepsilon_1$ which decreases for larger quasi-degenerate neutrino masses.
If extra information on how
$\mb{m}_\nu$ decomposes as the sum of $\mb{m}_T$ and $\mb{m}_H$ is available
(as e.g.\ happens when considering particular neutrino mass models)
the bound\eq{epsmax} can be strengthened, by replacing its last factor with
$\min (\sum_i m_{\nu_i}^2, \tilde{m}_H^2)$ where
$\tilde{m}_H^2\equiv {\rm Tr} \,{\mb{m}}_H^\dagger \mb{m}_H^{\phantom{\dagger}}$.

 When all terms in the effective Lagrangian are perturbative, 
 this CP asymmetry is smaller
 (often much smaller) than the generic absolute maximal value
 allowed by unitarity (i.e.\ by demanding positivity of all decay widths in eq.\eq{GammaTs}):
\bea 
\label{unit}
|\varepsilon_L|<2\min(B_L,B_H).
\eea
Complex soft terms in  supersymmetric triplet models
are a concrete example of an extra source
of CP-asymmetry that (unlike the neutrino mass contribution)
is not suppressed
for small $M_T$~\cite{DHHRR,SOFTN}.
Alternatively one can add extra terms to the triplet Lagrangian,
obtaining more complex phases.

\bigskip

\begin{figure}[t]
$$\hspace{-5mm}
\includegraphics[width=0.9\textwidth]{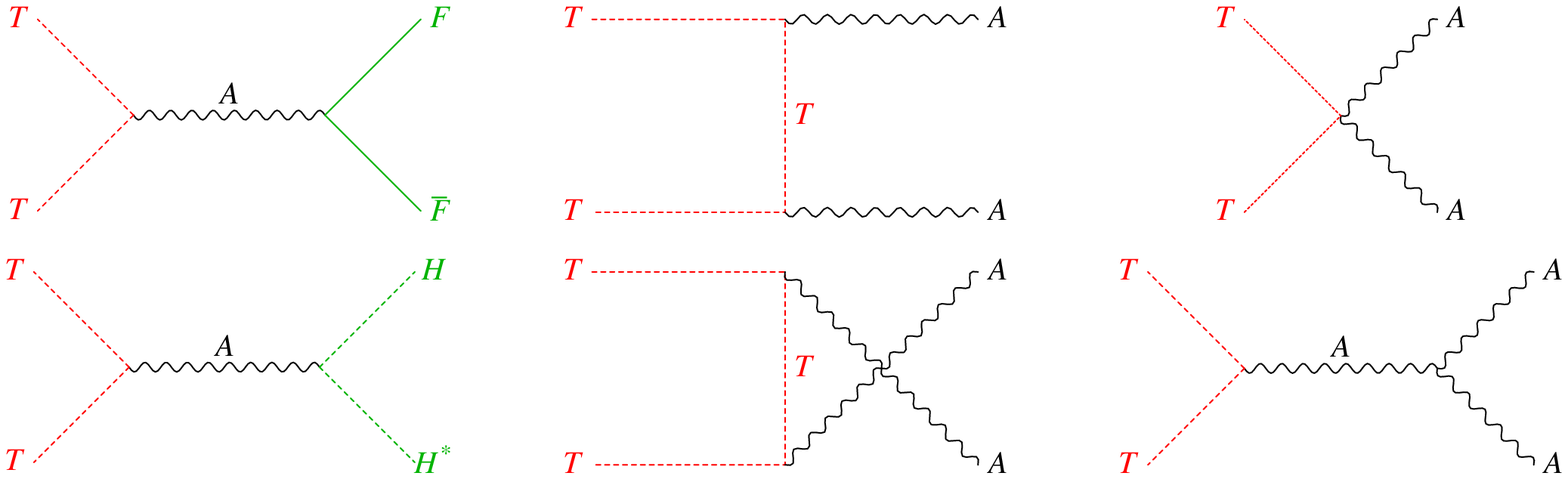}$$
\caption{\label{fig:FeynA}\em
Feynman diagrams that contribute to the interaction rate $\gamma_A$.}
\end{figure}

\subsection{Boltzmann equations}
We denote with $n_p$ the number density of the type `$p$' particles.
Boltzmann equations describe the evolution as function of $z\equiv M_T/T$ of the total $T,\bar{T}$ density
$\Sigma_T=(n_T + n_{\bar T})/s$ 
and of the asymmetries $\Delta_p = (n_p-n_{\bar p})/s$ stored 
in $p=\{T,L,H\}$.
They are: 
  \begin{eqnsystem}{sys:Boltz}
sHz \frac{d\Sigma_{T}}{dz} &=&
  -\bigg(\frac{\Sigma_{T}}{\Sigma_{T}^{\rm eq}}-1\bigg)\gamma_D
  -2\bigg(\frac{\Sigma_{T}^2}{\Sigma_{T}^{2\rm eq}}-1\bigg)\gamma_A \,, 
\label{eq:BoltzST}\\
sHz \frac{d\Delta_L}{dz} &=&  X
-2\gamma_DB_L(\frac{\Delta_L}{Y_L^{\rm eq}}+\frac{\Delta_T}{\Sigma_T^{\rm eq}}) ,
\label{eq:BoltzdL}\\
sHz \frac{d\Delta_H}{dz} &=& X
-2\gamma_DB_H(\frac{\Delta_H}{Y_H^{\rm eq}}-\frac{\Delta_T}{\Sigma_T^{\rm eq}}),
\label{eq:BoltzdH}
\\
sHz \frac{d\Delta_T}{dz} &=&-\gamma_D\left(\frac{\Delta_T}{\Sigma_{T}^{\rm eq}}+B_L
\frac{\Delta_L}{Y_{L}^{\rm eq}}-B_H \frac{\Delta_H}{Y_{H}^{\rm eq}}\right) ,
 \label{eq:BoltzdT}
\end{eqnsystem}
where $H$ is the Hubble constant at temperature $T$, 
$s$ is the total entropy density,
$Y_X=n_X/s$,
a suffix $^{\rm eq}$ denotes equilibrium values, $\gamma_P$ is the 
space-time density of type `$P$' 
processes computed in thermal equilibrium, and
\beq
\label{eq:Xdef}
X= \gamma_D \varepsilon_L \bigg(\frac{\Sigma_{T}}{\Sigma_{T}^{\rm eq}}-1\bigg)  -2
(\frac{\Delta_L}{Y_L^{\rm eq}} + \frac{\Delta_H}{Y_H^{\rm eq}}) 
(\gamma_{Ts}^{\rm sub}+\gamma_{Tt}).
\eeq
Notice that because of hypercharge (or electric charge) conservation 
only three out of the four Boltzmann equations are independent.
There exists a sum rule $$2 \Delta_T+\Delta_H-\Delta_L=0,$$
satisfied by eq.s~(\ref{sys:Boltz}).
An important comment to be made here is that,  since
triplets are not self-conjugated (unlike right-handed neutrinos),
there is a Boltzmann equation (\ref{eq:BoltzdT}) for $\Delta_T.$
As we discuss later, this structure of Boltzmann equations allows
new effects which are absent in the heavy neutrino leptogenesis.

The relevant processes contributing to triplet leptogenesis are:
\begin{itemize} 
\item Decays and inverse decays. $\gamma_D$ is the total decay 
space-time density of $T$ plus $\bar T$ decays, given by
the usual expression:
\beq \gamma_D =s\Gamma_T \Sigma_T^{\rm eq}  {\rm K}_1(z)/{\rm K_2}(z),
\eeq
where ${\rm K}_{1,2}$ are Bessel functions.

\item $\Delta T=2$ scatterings.
$\gamma_A$ is the space-time density of the $\SU(2)_L\otimes{\rm U}(1)_Y$
gauge scatterings $T\bar{T}\to \hbox{SM particles}$
shown in fig.\fig{FeynA}. This is a new effect not present with 
right-handed neutrinos.
The contributions of the various final states to the reduced cross 
section\footnote{We remind that reduced cross sections for $2\to  2$ 
scatterings are defined as
$\hat{\sigma} = \sum \int dt\,{|A|^2}/{8\pi s}$
where here $s,t$ are the usual Mandelstam variables
and  the sum runs over
initial and final spins and gauge indices.
The reaction densities are obtained as
$$\gamma=
\frac{T}{64 \pi^4} \int_{s_{\rm min}}^{\infty} ds~ s^{1/2}
 {\rm K}_1\bigg(\frac{\sqrt{s}}{T}\bigg)
  \hat{\sigma}(s).$$}
  $\hat\sigma_A$  are
\begin{eqnsystem}{sys:sigmaA}
\hat\sigma_A(T\bar{T}\to F\bar{F})&=&\frac{6g_2^4+5 g_Y^4}{2\pi } r^3,\\
\hat\sigma_A(T\bar{T}\to H\bar H)&=&\frac{g_2^4+g_Y^4/2}{8\pi} r^3,\\
\hat\sigma_A(T\bar {T}\to W^a W^b) &=& \frac{g_2^4}{\pi}\bigg[r(5+34/x)-
\frac{24}{x^2}(x-1)\ln\frac{1+r}{1-r}\bigg],\\
\hat\sigma_A(T\bar {T}\to YY,W^a Y) &=&
\frac{3g_Y^2 (g_Y^2 + 4 g_2^2)}{2\pi}\bigg[r(1+4/x)-\frac{4}{x^2}(x-2)\ln\frac{1+r}{1-r}\bigg],
\end{eqnsystem}
where $F$ denotes SM fermions, $r = \sqrt{1-4/x}$ and $x=s/M_T^2$.
The low-energy behavior ($\hat\sigma_A\propto r^3$ in the first two cases and
$\hat\sigma_A\propto r$ in the last two cases) is dictated by conservation 
of angular momentum.
Notice that by summing over $\SU(2)_L$ indices we include
all  `coannihilation' processes among different $X$ components.

\item  $\Delta L=2$ scatterings. 
Unlike in the singlet fermion case, the scalar triplet generates
 the $LL\leftrightarrow \bar H \bar H $ density rate
$\gamma_{Ts}$  only
by $s$-channel exchange;
and generates the $LH \leftrightarrow\bar L\bar H$ density rate $\gamma_{Tt}$ 
only by  $t$-channel exchange.
The reduced cross sections are
\begin{eqnsystem}{sys:DeltaL2}
 \hat\sigma_{Ts}&=&\frac{3xM_T^2}{4\pi v^4}\bigg[\frac{M_T}{1-x}+m_H\bigg]^2,\\
\hat\sigma_{Tt}&=&\frac{3 M_T^2}{4\pi v^4}\bigg[m_H^2 x + 4 m_H M_T
\bigg(1-\frac{\ln (1+x)}{x}\bigg)+
2M_T^2\bigg(-\frac{1}{1+x} + \frac{\ln (1+x)}{x}\bigg)\bigg]\!.\end{eqnsystem}
Note that in $\hat\sigma_{Ts}$ we included an extra factor 2 by hand which we 
took out in the coefficient of $\hat\sigma_{Ts}$ in eq.~(\ref{eq:Xdef}), 
so that $\hat\sigma_{Ts}$  becomes equal to $\hat\sigma_{Tt}$ in the low 
energy limit, where neutrino masses encode all $L$-violating effects:
\beq \hat\sigma_{Ts,Tt}\stackrel{s\ll M_T^2}{\simeq} \frac{m_\nu^2}{v^2} 
\frac{3s}{4\pi}.\eeq

In Boltzmann equations one must subtract from $\gamma_{Ts}$ the contribution
due to on-shell $T$-exchange, already taken into account by successive
decays and inverse-decays.
The subtracted reaction density is 
$\gamma_{Ts}^{\rm sub}\equiv \gamma_{Ts} - B_L B_H \gamma_D$
and can be more conveniently computed by replacing the $T$ propagator
with its off-shell part, as described in~\cite{GNRRS}.
Once this is done, $\gamma_{Ts}^{\rm sub},\gamma_{Tt}\ll \gamma_D$ at $T\sim M_T$,
unless the couplings $\lambda_{L,H}$ are big enough
that $\gamma_{Ts}^{\rm sub},\gamma_{Tt}$ are sizable, producing a 
strong wash-out of the baryon asymmetry. 
\end{itemize}
To conclude, we comment on various small additional effects.
We neglect all processes that give corrections of relative order 
$\alpha \circa{<} \hbox{few }\%$. 
We included RGE corrections to gauge couplings and neutrino masses: 
the result roughly is
$m_\nu(\hbox{High scale})\sim (1.2\div1.3) m_\nu$~\cite{bcst,GNRRS}.
We assumed that the lepton asymmetry is concentrated in a single flavour
(this e.g.\ typically happens if $\mb{m}_T\approx \mb{m}_\nu$ and neutrinos 
are hierarchical):
to fully include flavor one needs to evolve a $3\times 3 $ density matrix 
of lepton asymmetries
as described in~\cite{bcst}. 
At $T\circa{<}10^{11}\GeV$
sphalerons and SM Yukawa couplings
redistribute the asymmetries to left-handed quarks and
right-handed fermions respectively.
These effects can be taken into account
inserting appropriate ${\cal O}(1)$ redistribution factors (see e.g.~\cite{bcst});
apart from generating the baryon asymmetry of eq.\eq{uusm}, 
they have small impact on the dynamics of leptogenesis.

\section{Results for triplet scalar leptogenesis}\label{res}
%

%

The efficiency $\eta$ depends on 3 parameters
which can be chosen to be  $M_T$, 
$\lambda_L^2 \equiv \hbox{Tr}({\mb{\lambda}_L \mb{\lambda}_L^\dagger})$ and 
 $\tilde{m}_T$,
the contribution to neutrino masses mediated by triplet exchange.
We choose $\lambda_L$ because  in supersymmetric models it 
controls the renormalization induced Lepton Flavour Violating (LFV)
signals due to triplet interactions.
If $\mb{m}_T$ dominates neutrino masses, these signals are of
Minimal Flavour Violation~\cite{MFV} type;
if furthermore neutrinos are hierarchical then 
$\tilde{m}_T = (\Delta m^2_{\rm atm})^{1/2}\equiv m_{\rm atm}\approx 0.05\eV$.

Boltzmann equations for leptogenesis in  scalar triplet decays
 show two main 
qualitative differences with respect to the well known case of
right-handed neutrinos.
Firstly, gauge scatterings keep the triplet abundancy $\Sigma_T$ close 
to thermal equilibrium such that the final lepton asymmetry does not depend on
 the initial conditions. 
This effect is present also in the case of leptogenesis from Majorana fermion triplets~\cite{HLNPS},
and tends to reduce the efficiency.
It affects the efficiency in a negligible way only if the decay rate is much 
faster than the expansion rate because in this case the triplets decay 
before annihilating.
Secondly, unlike for Majorana triplets, due to the fact that the scalar 
triplets
have two independent types of
decay, and due to the related fact that there is one more Boltzmann
equation, the wash-out from decay can be avoided even if the decay is
much faster than the expansion rate. 
Lepton number is violated by the contemporaneous presence of
$\lambda_L$ and $\lambda_H$, so that the
 lepton asymmetry is 
washed-out  only when both
partial decay rates to leptons and Higgses are faster than the expansion rate.
Otherwise, even for a fast total decay
rate, a quasi maximal
efficiency can be obtained in large portions of the parameter space.


\begin{figure}[t]
$$\includegraphics[height=0.45\textwidth]{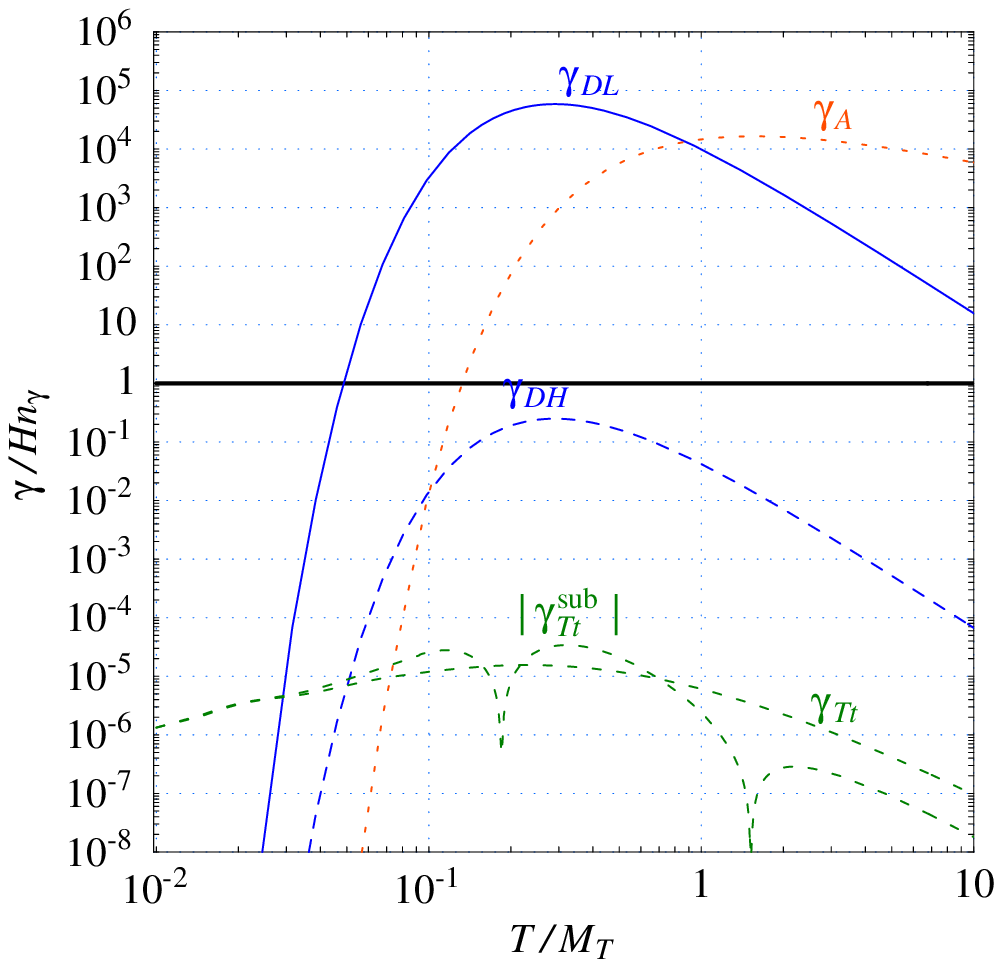}\hspace{0.1\textwidth}
\includegraphics[height =0.45\textwidth]{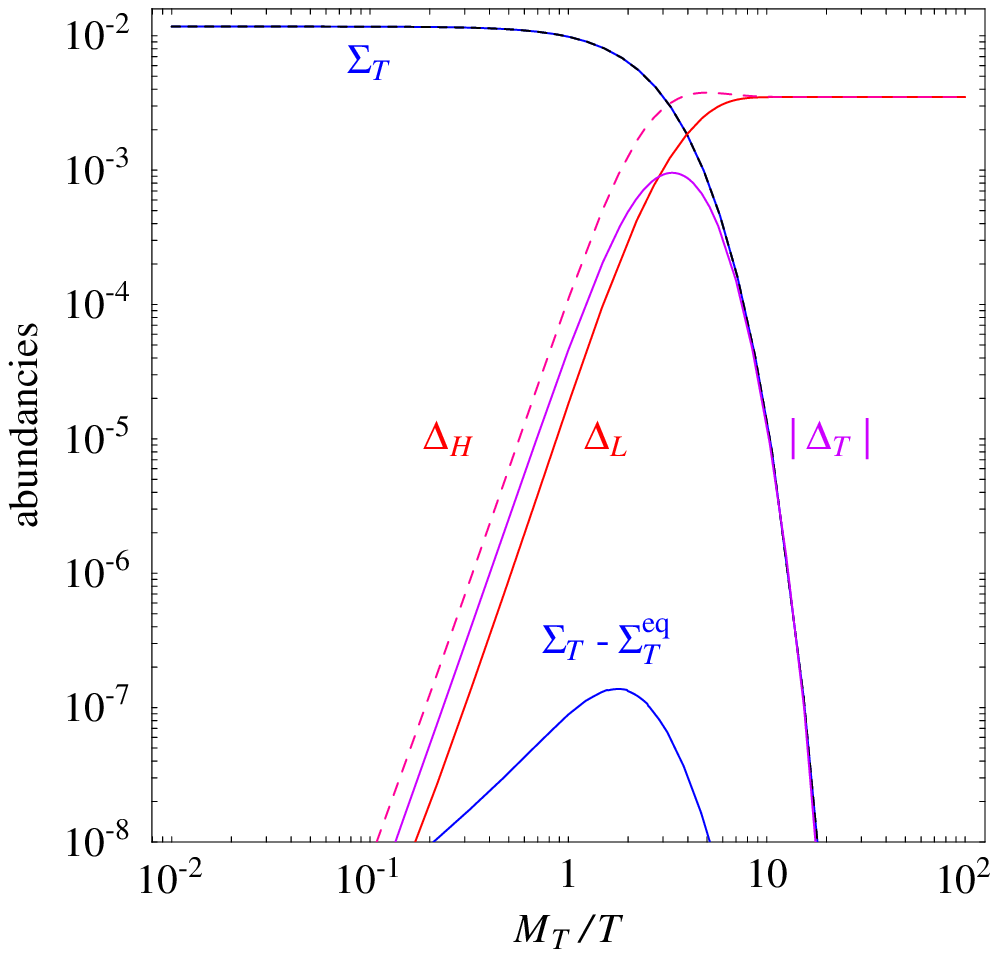}$$
\caption{\label{fig:gammaT}\em
Left panel:
interaction rates $\gamma_{DL},\gamma_{DH},\gamma_A,|\gamma_{Ts}^{\rm sub}|,
\gamma_{Tt}$ in units of $Hn_\gamma$: a value below 
${\cal O}(1)$ is `slow' on cosmological time-scale.
Right panel: 
 evolution of $\Sigma_T$ (indistinguishable from $\Sigma_T^{\rm eq}$,
 black dashed line),
$\Sigma_T-\Sigma_T^{\rm eq}$, $\Delta_L$, $\Delta_H$, $\Delta_T$.
Asymmetries are plotted in units of $\varepsilon_L$.
We take $M_1 = 10^{10}\GeV$, $\tilde{m}_T = (\Delta m^2_{\rm atm})^{1/2}\approx
0.05\eV$, $\lambda_L=0.1$ giving $\lambda_H\approx 2\cdot 10^{-4}$ and
efficiency  $\eta\approx 0.38$.}
\end{figure}

To demonstrate how a large efficiency $\eta\sim 1$ arises, we plot
in fig.\fig{gammaT} the interaction rates as function of temperature 
(left panel) as well as the evolution of the abundances of  $\Sigma_T$,
$\Sigma_T-\Sigma_T^{\rm eq}$, $\Delta_L$, $\Delta_H$, $\Delta_T$ (right panel)
for $M_T = 10^{10}\GeV$, $\tilde{m}_T = m_{\rm atm}$ and $\lambda_L=0.1.$ This choice of parameters implies 
$B_L\simeq 1\gg B_H\sim 10^{-5}$ and very different decay rates $\gamma_{DL} \equiv B_L\gamma_D$ and
$\gamma_{DH}\equiv B_H\gamma_D$. 
It happens that only $\gamma_{DL}$ is faster than the expansion rate, so that 
the washout of the produced $L$ asymmetry is not effective.
Furthermore, gauge scatterings do not give a significant suppression
despite being much faster than the expansion rate, 
because they are not much faster than $\gamma_{D}$ (see fig.\fig{gammaT}a):
triplets decay before annihilating. While gauge scatterings keep
triplets in equilibrium, all asymmetries  grow much larger 
than $\Sigma_T-\Sigma_T^{\rm eq}$. Altogether the resulting efficiency in this
example is quasi-maximal: $\eta\approx 0.38$.

\smallskip

\begin{figure}[t]
$$
\includegraphics[width=0.45\textwidth]{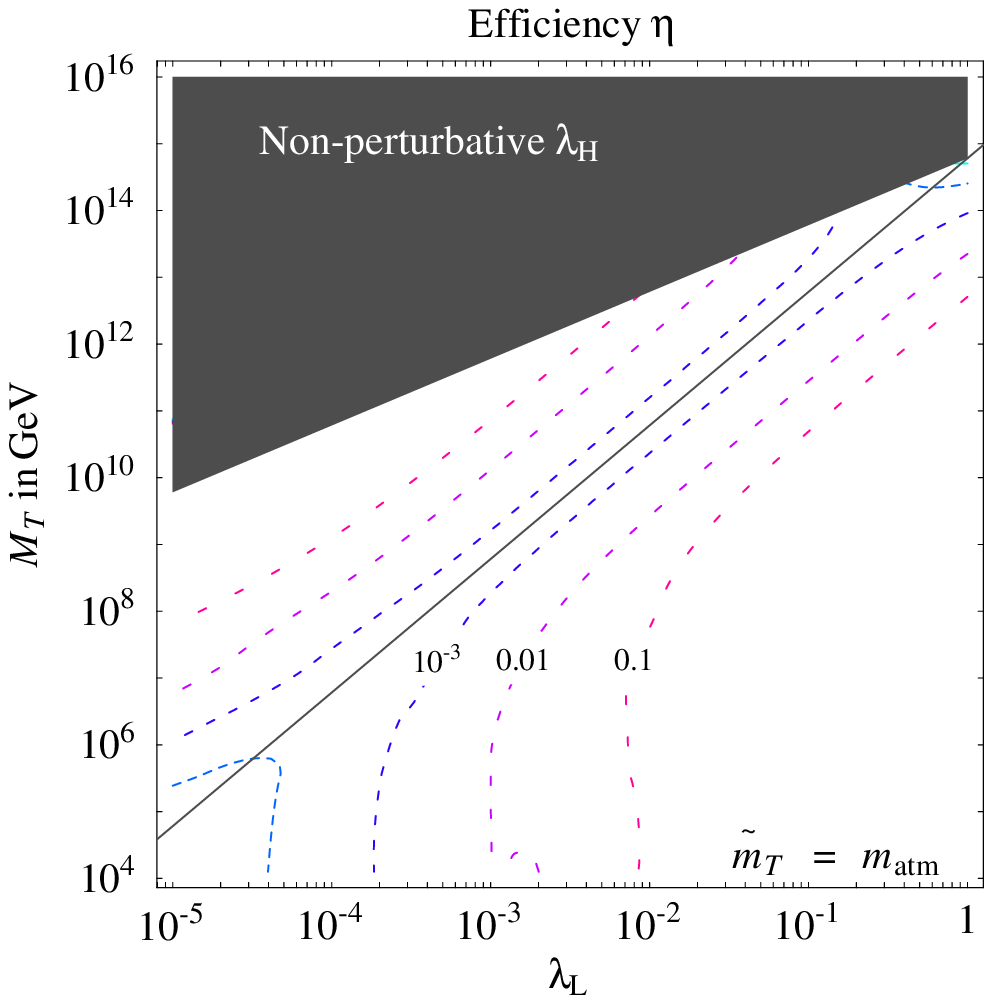}\qquad
\includegraphics[width=0.45\textwidth]{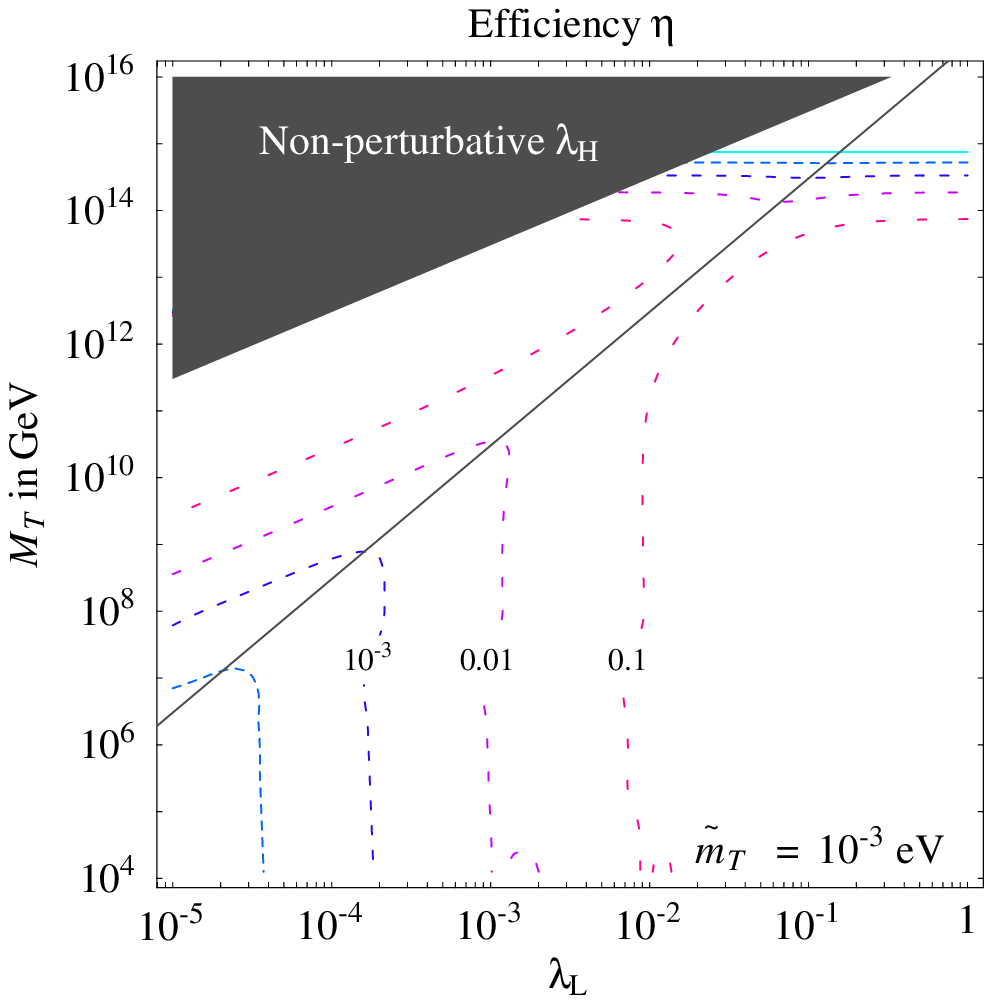}$$
\caption{\label{fig:eta}\em
Iso-curves of the efficiency $\eta$ in the ($\lambda_L,M_T$) plane
at fixed $\tilde{m}_T = 0.05\eV$ (left panel) and $\tilde{m}_T = 10^{-3}\eV$ (right panel).
The diagonal line corresponds to ${\rm BR}(T\to \bar L\bar L) = {\rm BR}(T\to HH)=1/2$
and shading covers regions with $\lambda_H>1$.}
\end{figure}

\begin{figure}[t]
$$
\includegraphics[width=0.45\textwidth]{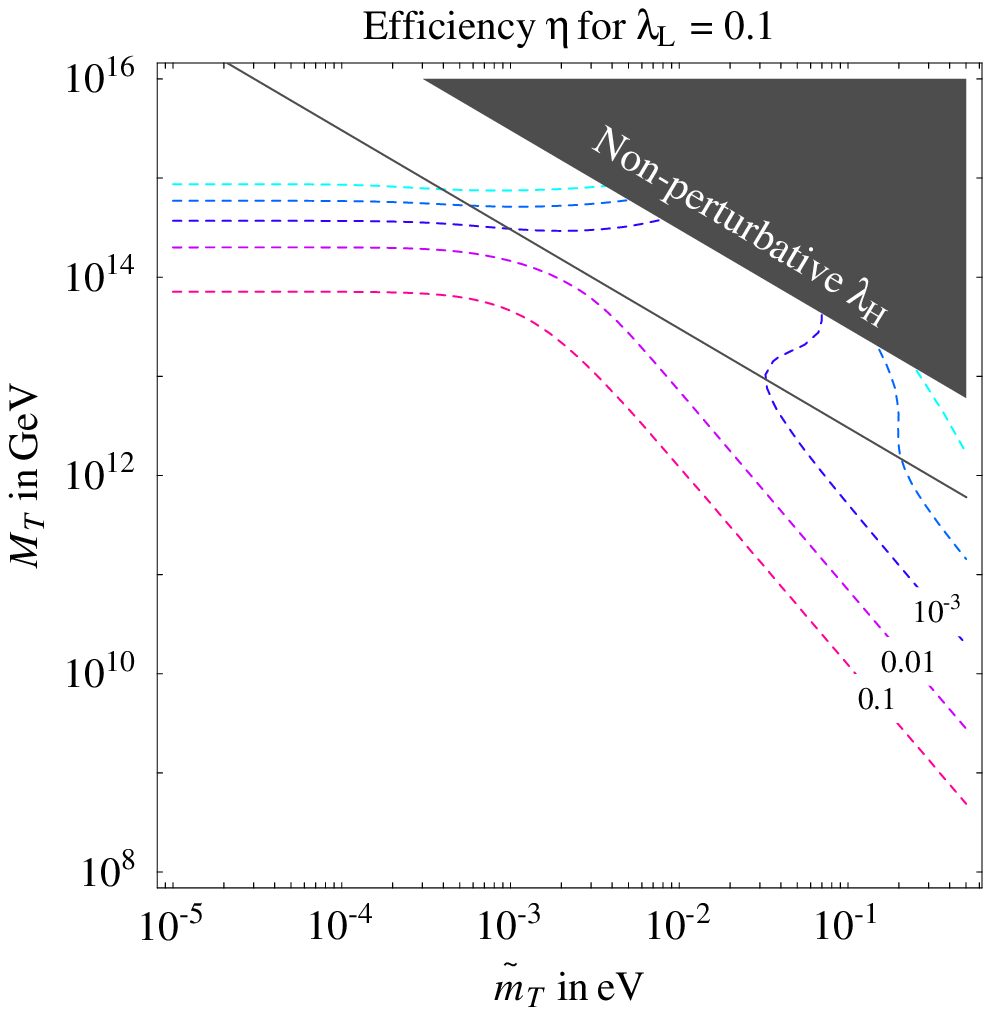}\qquad
\includegraphics[width=0.45\textwidth]{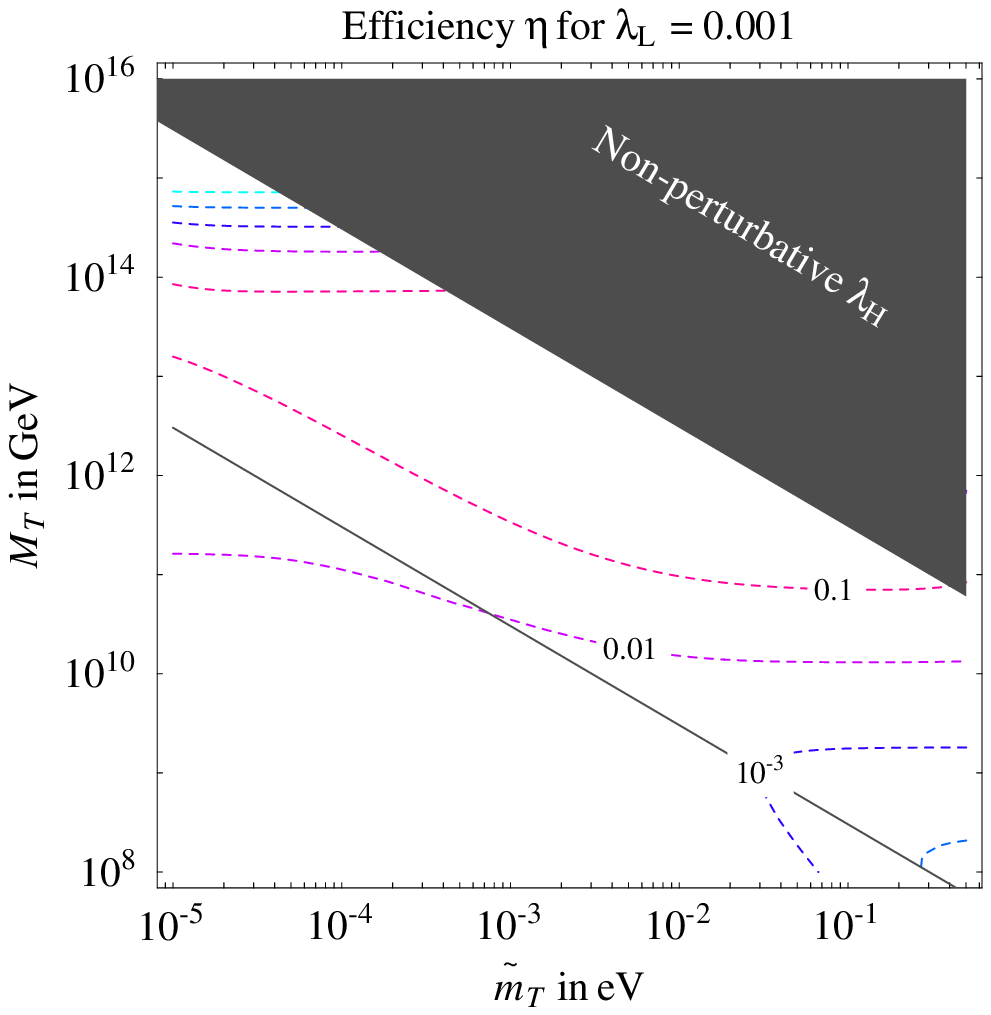}$$
\caption{\label{fig:EtamM}\em
Efficiency $\eta$  in the $(\tilde{m}_T,M_T)$
plane at fixed $\lambda_L=0.1$ (left panel) and  $\lambda_L=0.001$ (right panel).
The diagonal line corresponds to ${\rm BR}(T\to \bar L\bar L) = {\rm BR}(T\to HH)=1/2$
and shading covers regions with $\lambda_H>1$.}
\end{figure}

Let us present a more technical explanation of the behavior discussed above.
In general, Boltzmann equations
reduce to thermal equilibrium  (giving a vanishingly small efficiency $\eta$)
when the source terms in their
right-handed sides have coefficients much larger than the expansion rate.
From their explicit form one can see that when $B_L\to 0$ or when $B_H\to 0$ the combinations 
of $\Delta_L,\Delta_H,\Delta_T$
that are forced to vanish become linearly dependent, 
so that one combination gets not washed out and can store a large lepton asymmetry.
This behavior is possible because $B_L$ and $B_H$ can be  
different from each other, which results in the additional  
Boltzmann equation, eq.~(\ref{eq:BoltzdT}).\footnote{An aside 
comment: Boltzmann equations for decays of 
right-handed sneutrinos have a similar form
as leptogenesis from decays of scalar triplets.
In the sneutrino case the new effect discussed here is typically negligible, 
because unbroken supersymmetry forces equal branching ratios for the 2 different 
sneutrino decay modes.}
We can analytically explain the numerical results in fig.~3.
In this example
$\Gamma(\bar T\rightarrow LL) \gg  H$ and 
$\Gamma(T \rightarrow H H) \circa{<} H$.
Then: 
\begin{itemize}
\item[a)] Eq.~(\ref{eq:BoltzST}) is 
dominated by the $\gamma_D$ term which puts $\Sigma_T$ close to 
$\Sigma_T^{\rm eq}$; gauge scatterings have a negligible effect.
\item[b)] The washout terms in eq.~(\ref{eq:BoltzdH}) can be neglected
thanks to $B_H\ll1$, and a large $\Delta_H$ asymmetry develops:
it does not depend on $\gamma_D$, because it is
proportional to $\gamma_D$ times $\Sigma_T/\Sigma_T^{\rm eq}-1$ 
(which is inversely proportional to $\gamma_D$). 
Indeed the approximate analytical solution is 
$\Delta_H(T)\approx\varepsilon_L [\Sigma_T^{\rm eq}(T\gg M_T)- \Sigma_T^{\rm eq}(T)]$
where $\Sigma_T^{\rm eq}(T\gg M_T)$ is a constant and
$\Sigma_T^{\rm eq}(T\ll M_T)\simeq 0$.

\item[c)] Due to 
the $2 \Delta_T+\Delta_H-\Delta_L=0$ sum rule, 
an equally large $2 \Delta_T-\Delta_L$ asymmetry is also 
produced.
\item[d)] Subsequently, as all triplets decay, $\Delta_T$ goes 
to zero so that the large $2 \Delta_T-\Delta_L$ becomes a 
large $\Delta_L$ asymmetry, with efficiency of order unity.
\end{itemize}


To show the dependence of the efficiency in the 3 parameters $M_T$, 
$\lambda_L$ and $\tilde{m}_T$, various isocurve plots can be considered. 
Fig.\fig{eta} shows  $\eta$ in the ($\lambda_L,M_T$) plane for
$\tilde{m}_T  = m_{\rm atm}=0.05\eV$ (left panel) and for
$\tilde{m}_T = 10^{-3}\eV$ (right panel).
Fig.\fig{EtamM} shows $\eta$ in 
the  $(\tilde{m}_T,M_T)$ plane for  $\lambda_L=0.1$ (left panel) and 
 $\lambda_L=0.001$ (right panel). 
For $B_L=B_H=1/2$ (represented in the plot by the diagonal line) 
the efficiency
is dominantly determined by $\gamma_D$ and $\gamma_A$;
$\Delta L =2$
scatterings are relevant only above $\sim 10^{14}$~GeV).
By comparing their rates with the Hubble rate
\beq
\frac{\gamma_D}{H n_\gamma}\Big|_{T\approx M_T}\approx
\frac{\Gamma_T}{H}\Big|_{T\approx  M_T}\approx 
\frac{\tilde{m}_T}{10^{-3}\,\eV},\qquad
\frac{\gamma_A}{H n_\gamma}\Big|_{T\approx  M_T}\simeq 
\frac{10^{14} \GeV}{M_T}\,,
\eeq 
one can understand our numerical results:
for $\tilde{m}_T=0.05$~eV decays are more important 
than gauge scatterings and 
$\eta \approx 10^{-3}$ almost independently on $M_T$;
the reverse happens for $\tilde{m}_T=10^{-3}$~eV and $\eta$ decreases with $M_T$.
For $B_L\gg B_H$ ($B_H\gg B_L$), the efficiency is larger
thanks to $\gamma_{DL}<H n_\gamma$ ($\gamma_{DH}<H n_\gamma$).
In conclusion, numerical results shows the features anticipated above:
i.) The efficiency is maximal, $\eta\sim 1$, when either $B_L\gg B_H$ or $B_H\gg B_L$
and minimal when $B_L=B_H=1/2$.
ii.) Even when $B_L=B_H=1/2$ the efficiency does not depend much 
on $M_T$, especially for larger $\tilde{m}_T\circa{>}10^{-3}\eV$,
remaining relatively large even for $M_T\sim\TeV$.

\begin{figure}[t]
$$
\includegraphics[width=0.45\textwidth]{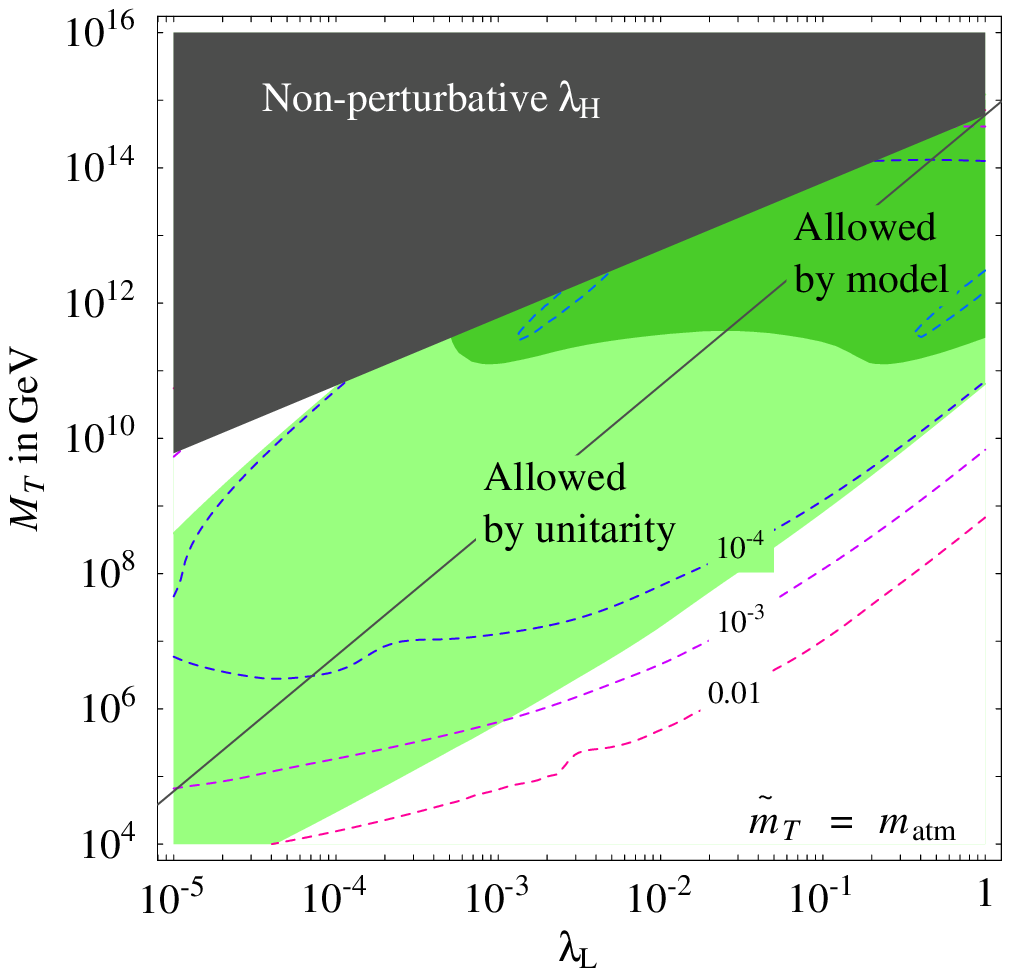}\qquad
\includegraphics[width=0.45\textwidth]{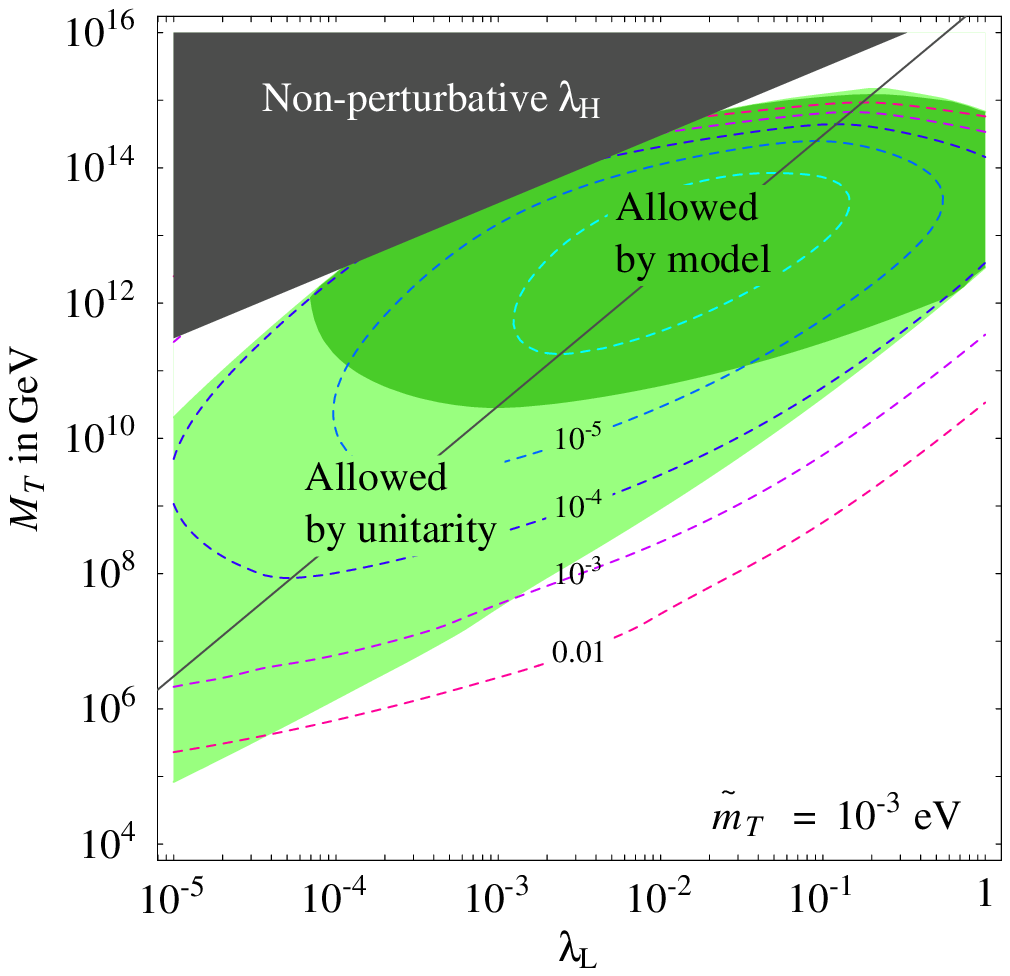}$$
\caption{\label{fig:isoepsL}\em
Iso-curves of value of $\varepsilon_L/\sqrt{4B_L B_H}$ needed to have successful leptogenesis 
in the $(\lambda_L,M_T)$
plane for two values of $\tilde{m}_T$ as indicated in the figure. The 
dark-green (grey) 
region is the allowed region assuming that the CP-asymmetry
arises from heavier sources of 
neutrino masses, i.e.~fulfilling eq.~(\ref{eq:epsmax}).
The light green (grey) region is obtained assuming that the CP-asymmetry is bounded only 
by unitarity, eq.~\rfn{unit}.}
\end{figure}

\medskip



\medskip

We now come to the calculation of the produced baryon asymmetry. The 
values of $\varepsilon_L$ necessary for obtaining
the observed baryon asymmetry,
$n_B/n_\gamma\approx 6.2 \cdot 10^{-10}$, can be obtained 
straightforwardly from fig.s~\ref{fig:eta}, \ref{fig:EtamM} 
using eq.~(\ref{eq:uusm}). 
Notice that, while 
the efficiency is maximal when $B_L\to 0$ or $B_H\to 0$,
in these limits the CP-asymmetry gets suppressed, 
in a way which depends on its origin.
In the minimal model $\varepsilon_L\propto \sqrt{B_L B_H}$, see eq.\eq{DIT}.
To illustrate this fact,
fig.\fig{isoepsL} shows the values of $|\varepsilon_L|/\sqrt{4 B_L B_H}$ needed for successful 
leptogenesis for 
 $\tilde{m}_T=m_{\rm atm}$ (left panel) and $\tilde{m}_T=10^{-3}$ eV (right 
panel). 
Using the upper bound  on $|\varepsilon_L|$ of eq.~(\ref{eq:epsmax})
(valid assuming the minimal model where $\varepsilon_L$ is generated only by other sources
of neutrino masses much heavier than the scalar triplet; we evaluate
it assuming hierarchical neutrino masses)
restricts the ranges of $M_T$ and $\lambda_L$ which can lead to successful 
leptogenesis to the region shaded in dark green. 
This allowed region covers a wide range of $\lambda_L$ values:
the decrease of the maximal 
$|\varepsilon_L|$ when  $B_L\to 0$ or $B_H\to 0$
is roughly compensated by the increase in the efficiency, such that
successful leptogenesis does not need $B_L\approx B_H$.
In this minimal model $\varepsilon_L\propto M_T$, so that
successful leptogenesis needs a heavy enough triplet.
Assuming a hierarchical spectrum of light neutrinos 
(i.e.~setting $\sum_i m_{\nu_i}^2=\Delta m^2_{\rm {atm}}$ 
in eq.~(\ref{eq:epsmax})) we find the model-independent bounds:
\begin{eqnsystem}{sys:bounds}
\label{eq:MTbound}
M_T & > & 2.8 \cdot 10^{10}\,\mrm{GeV} \quad (\tilde{m}_T=0.001\,\eV);\\
M_T & > & 1.3 \cdot 10^{11}\,\mrm{GeV} \quad (\tilde{m}_T=0.05\,\eV). 
\end{eqnsystem}
A stronger bound on the CP asymmetry holds in models where heavier sources of neutrino masses
are predicted to give a small contribution $\mb{m}_H$. 
A too small $\mb{m}_H$ prevents successful leptogenesis, and
the bounds of eqs.~(\ref{sys:bounds}) become stronger, by a factor
well approximated by $(\sum_i m_{\nu_i}^2)^{1/2}/\tilde{m}_H$ 
where $\tilde{m}_H^2\equiv {\rm Tr}\,\mb{m}_H^\dagger\mb{m}_H$.
For example, 
for $\tilde{m}_H = (\Delta m^2_{\rm sun})^{1/2}=0.007$~eV this gives:
$M_T  >  8 \cdot 10^{11}\,\mrm{GeV}$ (with $\tilde{m}_T=0.05\,\eV$).
These precise constraints can be compared to the estimated constraint: 
$M_T > 10^{11-12}$~GeV \cite{scaltriplepto,hs,seesaw25TH}.

For larger values of $\sum_i m_{\nu_i}^2$ the constraints of 
eqs.~(\ref{sys:bounds}) get relaxed by a
$(\Delta m^2_{\rm atm}/\sum_i m_{\nu_i}^2)^{1/2}$ factor: i.e.~by  one order of magnitude for quasi-degenerate neutrinos with $m_\nu\approx0.5\eV\approx 10 m_{\rm atm}$.
In fact, by increasing the neutrino mass scale keeping $m_T$ fixed,
the asymmetry increases but the efficiency remains unchanged~\cite{hs}.



\medskip

We expect that adding supersymmetry does not significantly affect our results.
More precisely, the interaction rates and the CP-asymmetry generated 
by neutrino masses
become ${\cal O}(2)$ times bigger,
and the numerical coefficient in eq.\eq{uusm} remains almost unchanged.
This means that the constraint on $M_T$ 
in eq.~(\ref{eq:MTbound})
conflicts with the gravitino
constraint on the maximal big-bang temperature~\cite{nucleo}.
From this point of view, triplet leptogenesis is not better than the other two
mechanisms that can mediate tree-level masses.
In all 3 cases this incompatibly can be circumvented in many ways;
in particular extra sources of CP-violation unrelated to neutrino masses
(and related e.g.\ to soft supersymmetry-breaking terms)
allow much larger 
asymmetries and the lower bound on $M_T$ can be considerably 
relaxed.
In general the CP-asymmetry is bounded only by eq.~(\ref{unit}),
that does not depend on $M_T$: 
a light triplet can produce successful leptogenesis
because its efficiency remains large enough.
Fig.\fig{isoepsL} shows that even $M_T\sim\TeV$ is allowed:
the region allowed  by the unitarity bound is shaded in light green.

Notice that the fact that relatively large values of $\lambda_L$ are 
compatible with thermal triplet leptogenesis, see fig.\fig{isoepsL},
has an interesting consequence
in SUSY models: 
RGE corrections imprint $\lambda_L$ in slepton masses.
If $\mb{\lambda}_L$ violates lepton flavour this effect induces LFV charged-lepton processes
with possibly detectable rates.
As usual, the predicted LFV rates depend also on sparticle masses which can
be measured at colliders. Taking into account naturalness considerations
and experimental bounds and hints, we give our numerical examples in the leading log approximation for
$m_0=M_{1/2}=200\GeV$, $A_0=0$, $\tan\beta=5$.
The most relevant effects are then  approximately given by
(see also~\cite{anna})
\begin{eqnsystem}{sys:BR}\label{eq:BR}
{\rm BR}(\mu\to e\gamma) &\approx& 1.2\cdot 10^{-4}r\, 
|\mb{\lambda}_L^\dagger\cdot \mb{\lambda}_L|_{e\mu}^2 
\ln^2\frac{M_{\rm GUT}}{M_T},\\
{\rm BR}(\tau\to\mu\gamma) &\approx& 2\cdot 10^{-5}r|\mb{\lambda}_L^\dagger \cdot \mb{\lambda}_L|_{\mu\tau}^2 
\ln^2\frac{M_{\rm GUT}}{M_T}
\end{eqnsystem}
where $r\approx (\tan\beta/5)^2(200\GeV/M_\mathrm{SUSY})^4$
  equals 1 at our reference point.
$\lambda_L^{\tau\mu}\circa{>}10^{-1}$ gives rise to detectable $\tau\to \mu\gamma$ rates,
$\lambda_L^{e\mu}\circa{>}10^{-2}$ gives rise to detectable $\mu\to e\gamma$ rates.

\bigskip


\section{Conclusions}\label{concl}
We computed the efficiency factor $\eta$ that summarizes the dynamics of scalar triplet thermal leptogenesis.
Despite the presence of gauge interactions, that tend to maintain the triplet abundancy
very close to thermal equilibrium, one can have even maximal efficiency, $\eta\sim 1$, for any triplet mass $M_T$,
even $M_T\sim \TeV$.
This happens when 
i) one of the two decay rates ($T\to \bar L\bar L$ or $T\to HH$)
is faster than the annihilation rate;
ii)  the other one is slower than the expansion rate.
Thanks to i) gauge scatterings are ineffective: triplets decay before annihilating.
Thanks to ii) fast decays do not produce
a strong washout of the lepton asymmetry (and consequently a small efficiency $\eta$),
because  lepton number is violated only by the contemporaneous presence
of the two $T\to \bar L\bar L$ and $T\to HH$ processes.
Our numerical results are obtained by writing and solving the 
full set of Boltzmann equations, eq.s~(\ref{sys:Boltz}).

 \medskip

We obtained in eq.\eq{DIT} an expression for the CP asymmetry in triplet decays, assuming
 that it is related to neutrino masses.
 Neutrino masses   $\mb{m}_\nu = \mb{m}_T + \mb{m}_H$ can be written
 as the sum of the
triplet contribution  $\mb{m}_T$,
plus the contribution $\mb{m}_H$ from any other sources.
The suffix $H$ indicates that we assume that other sources are  much $H$eavier than $M_T$,
 such that at energies $E\circa{<}M_T$
 all their effects are encoded in $\mb{m}_H$.
This assumption allows us to derive an upper bound on the 
triplet CP-asymmetry,
eq.\eq{epsmax},
analogous (but not equal) to the bound that holds in the
right-handed neutrino case.

\medskip

Combining $\eta$ with the maximal CP-asymmetry generated by neutrino masses
allows us to derive
a lower bound on the triplet mass $M_T$ which 
varies between $10^9$~GeV and $10^{12}$~GeV, depending on the neutrino 
mass contribution of both triplet and heavier source of neutrino mass, 
see eq.s~(\ref{sys:bounds}).
This also leads on lower and upper bounds on the
triplet Yukawa coupling to leptons, that in supersymmetric models
induces LFV processes such as $\mu\to e\gamma$ and $\tau\to\mu\gamma$.
Neutrino masses depend on the product $\lambda_L\lambda_H$ of
triplet couplings to leptons and Higgs.
Leptogenesis separately depends on $\lambda_L$ and $\lambda_H$, but
adds no more information: 
the region $\lambda_L\sim \lambda_H$ is not singlet out,
due to the unexpected  behavior of the efficiency.
As a result large values of $\lambda_L$, and therefore 
large rates of LFV 
processes, are allowed.

\medskip

By relaxing the assumption on the origin of the CP-asymmetry,
it can reach larger values (bounded only by unitarity) that can be
realized e.g.\ in  supersymmetric models with complex soft terms (`soft leptogenesis'),
especially if the triplet mass is not much heavier than the scale of SUSY breaking.
In this context the competing effect of the CP-asymmetry and of the efficiency
favors $\lambda_L\sim\lambda_H$, allowing
successful triplet thermal leptogenesis even at $M_T\sim\TeV$
provided that $\lambda_L/\lambda_H\circa{<}(0.1\div 10)$.

\paragraph{Acknowledgments}
We thank S. Davidson for useful discussions.
We thank the CERN TH division, where this work was partly produced.
The work of T.H.~is supported by the EU Marie Curie fellowship HPMF-CT-01765.
The work of A.S.\ is supported in part by the European Programme `The Quest For Unification', contract MRTN-CT-2004-503369.
M.R. is supported by ESF grant nr.\ 6140.
   
\appendix
 
\footnotesize
\begin{multicols}{2}
  
\end{multicols}
\end{document}